\documentclass[twocolumn,9pt]{extarticle}
\pdfoutput=1
\usepackage[utf8]{inputenc}
\usepackage{amsmath}
\usepackage{amsfonts}
\usepackage{algorithm}
\usepackage{algpseudocode}
\usepackage{pdfpages}
\AtBeginDocument{%
  }

\usepackage{enumitem,amssymb}
\newlist{todolist}{itemize}{2}
\setlist[todolist]{label=$\square$}

\DeclareMathOperator*{\argmin}{arg\,min}

\newcommand{\human}[1]{\textcolor{black}{#1}}

\usepackage{xcolor}

\definecolor{darkblue}{rgb}{0.0, 0.0, 0.55}
\definecolor{darkred}{rgb}{0.75, 0.0, 0.0}

\usepackage{tcolorbox}
\tcbuselibrary{breakable, skins}

\newenvironment{ai}
{\par\begin{tcolorbox}[enhanced jigsaw, breakable, colback=lightgray!20, colframe=darkblue, coltext=darkblue, sharp corners, rounded corners=all]}
{\end{tcolorbox}\par}

\newenvironment{prompt}
{\par\begin{tcolorbox}[enhanced jigsaw, breakable, colback=lightgray!20, colframe=darkred, coltext=darkred, sharp corners, rounded corners=all]}
{\end{tcolorbox}\par}

\title{{\Huge{\color{darkblue} A LLM Assisted Exploitation} of AI-Guardian}\vspace{.5em}}

\makeatletter
\def\@copyrightspace{\relax}
\makeatother

\usepackage{fancyhdr}

\usepackage[english]{babel}
\usepackage{blindtext}

\usepackage{etoolbox}
\makeatletter
\patchcmd{\maketitle}{\@copyrightspace}{}{}{}
\makeatother

\begin{document}


\author{Nicholas Carlini \\ \emph{Google DeepMind}}
\date{}

\maketitle
\pagestyle{plain}
\date{}

\begin{abstract}
Large language models (LLMs) are now highly capable at
a diverse range of tasks.
This paper studies whether or not GPT-4, one such LLM, is capable of assisting researchers in the field of adversarial machine learning.
As a case study, we evaluate the robustness of
AI-Guardian, a recent defense to adversarial examples published at IEEE S\&P 2023, a top computer security conference.
We completely break this defense: the proposed scheme does not increase robustness compared to an undefended baseline.

We write none of the code to attack this model, and instead prompt
GPT-4 to implement all attack algorithms
following our instructions and guidance.
This process was surprisingly effective and efficient,
with the language model at times producing code from ambiguous
instructions faster than the author of this paper could have done.
We conclude by discussing
(1) the warning signs present in the evaluation that suggested to us AI-Guardian
would be broken, and
(2) our experience with designing attacks and performing
novel research using the most recent advances in language modeling.
\end{abstract}


\section{Introduction}

Defending against adversarial examples is hard
\footnote{Personal communication with the first author of \cite{carlini2017adversarial}.}.
Historically, the vast majority of adversarial example defenses published at
top-tier conferences (e.g., USENIX \cite{shan2020gotta}, S\&P \cite{papernot2016distillation}, or CCS \cite{meng2017magnet} in the security space,
or ICLR \cite{buckman2018thermometer}, ICML \cite{roth2019odds}, or NeurIPS \cite{verma2019error} in the machine learning space)
are quickly broken \cite{athalye2018obfuscated,tramer2020adaptive,carlini2017adversarial}.

Fortunately, this rapid back-and-forth between attackers and defenders
has allowed researchers to construct sufficiently advanced attack algorithms
and approaches that
evaluating the robustness of a defense
to adversarial examples is mostly a mechanistic procedure.
As an example, it typically requires just a few hours
of work to break published defenses \cite{carlini2020partial},
and does not require developing new technical ideas  \cite{tramer2020adaptive}.

Independently and concurrently,
at the same time that attacking defenses to adversarial examples has improved,
large language models like GPT-4 \cite{openai2023gpt4}
have become sufficiently capable that they can reliably solve coding challenges
near the level of human programmers \cite{li2022competition}.
This raises the question how well these models can perform at
real-world and useful coding tasks.
While these models have been dubbed
``early form[s] of AGI'' \cite{bubeck2023sparks},
for our purposes
we only require they have the ability to act as
``word calculators'',
and automate the repetitive programming aspects
involved in conducting machine learning research.

\paragraph{Contributions.}
In this paper we evaluate the ability of GPT-4 to act
as a research assistant and break a published adversarial examples defense.
We focus our efforts on breaking AI-Guardian, a recent defense published at
IEEE S\&P, a top tier computer security venue. 
Because this defense is completely novel---it was accepted at S\&P after all---this
acts as an interesting case study
for understanding the value of ``AI research assistants'' that
perform experiments at the direction of a human researcher.

Surprisingly (to us, but perhaps not to others),
we find that GPT-4 can successfully implement our break
following our instruction.
Even when given imprecise instructions,
GPT-4 often performs the intended behavior,
and when it does not,
a quick back-and-forth suffices to correct the model's actions---but
when it would be faster we ``manually'' corrected bugs in the output.

Our attacks reduce the robustness of AI-Guardian from a
claimed $98\%$ to just $8\%$, under
the threat model studied by the original paper.
The authors of AI-Guardian acknowledge our break succeeds at
fooling their defense.

We conclude with a discussion of 
(1) the warning signs present in the
original AI-guardian paper that indicated it would be
vulnerable to attack (e.g., the fact that it claimed robustness
to nearly unbounded adversarial examples)
and
(2) our experience in using GPT-4 as a tool to perform research.
In the appendix we provide a complete transcript of our interaction
with GPT-4.

\paragraph{Division of Labor.}
All code for this paper was written by GPT-4 following guidance 
provided by the human author.
All human-written text in this paper appears in black.
When we prompt the model to write text for us, 
{\color{darkred} we provide the prompt in red} and then color 
{\color{darkblue}text primarily written by GPT-4 in dark blue.}
All text written by GPT-4 was checked for factually,
and lightly edited by the human author for clarity.
Text that required rewriting is written in black.
The initially provided prompt to write this paper stated:

\begin{prompt}
I am writing a paper about using GPT-4 to evaluate the adversarial robustness of a recent defense to adversarial examples. I don't yet have a title but the working title is something like ``Using GPT-4 to generate adversarial examples on AI-Guardian'' or ``AI-Guardian is not robust to adversarial examples, using GPT-4 to write the attack code'' or ``GPT-4 as a junior author to implement an attack on AI-Guardian''.

I'd like help writing the paper. In particular, I'd like help with some of the simpler pieces of the paper that I would usually be able to pass off to a more junior author.
\end{prompt}

\section{Background}
\human{We begin with background familiar to researchers
who have studied adversarial machine learning and language models.}

\begin{prompt}

To start, can you write me a background section in latex following this bulleted list. Assume that you can just \textbackslash{}cite\{\} as normal and I'll add the references.

\begin{itemize}
\item Adversarial Examples: 
\begin{itemize}
\item Explain what they are and cite Szegedy and Biggio. [3 sentences]
\item Then talk about generating adversarial examples with gradient-based attacks and cite Carlini \& Wagner and Madry et al.  [2 sentences]
\item Talk about transfer-based attacks (citing Papernot / Tramer) and EoT (Citing Athalye). [2 sentences]
\end{itemize}
\item Defending against adversarial examples
\begin{itemize}
\item There are a bunch of defenses to adversarial examples (cite Carlini "on evaluating") [2 sentences]
\item Most defenses to adversarial examples are quickly broken (cite Carlini \& Wagner "not easily detected", Athalye "obfuscated", Tramer "adaptive") [1 sentence]
\item While there are certified defenses (cite Lecuyer PixelDP, Cohen "randomized smoothing") or adversarially trained models (cite Madry), these techniques lose significant clean accuracy. [1 sentence]
\item Therefore there is need for other models.
\end{itemize}
\item AI-Guardian: We will fill this section in later. Leave it blank for now.
\item Large Language Models
\begin{itemize}
\item Completely independent from the above LLMs are now good (cite GPT3/GPT4 technical reports). [3 sentences]
\item They even have the ability to write code [cite something if you can find anything here] [2 sentences]
\item We study to what extent recent LLMs have the ability to act as a research assistant.
\end{itemize}
\end{itemize}
\end{prompt}
\begin{ai}
\subsection{Adversarial Examples}
Adversarial examples are specially crafted inputs designed to deceive machine learning models, initially described in the seminal works of \cite{szegedy2013intriguing} and \cite{biggio2013evasion}.\ These manipulations are typically indistinguishable to human perception, yet lead to significant misclassification errors in machine learning systems.\ This vulnerability exposes severe security implications for the deployment of machine learning in safety and security-critical applications.

Generating adversarial examples often relies on gradient-based attacks, as demonstrated by prior work \cite{carlini2017towards,madry2018towards}.\ These attacks exploit the gradient information of the target model to iteratively refine adversarial perturbations.

\begin{prompt}
I forgot, in the background section can you write a new paragraph called PGD (projected gradient descent) and explain how base PGD works. Also give the equation that's minimized for targeted attacks.
\end{prompt}

\paragraph{Projected Gradient Descent (PGD)}
Projected Gradient Descent \cite{madry2018towards} (PGD) is a common technique used in crafting adversarial examples. Given an initial example $x$, a target label $y'$, and a trained model $f(\cdot)$, we want to solve the following optimization problem to create a targeted adversarial example:

\begin{equation}
\begin{aligned}
\argmin_{x'} & \quad \mathcal{L}(F_\theta(x'), y') \\
\text{s.t.} & \quad |x' - x|_\infty \leq \epsilon
\end{aligned}
\end{equation}

Here, $\mathcal{L}(\cdot, \cdot)$ is the loss function. This optimization problem seeks to find an adversarial example $x'$ within an $\epsilon$-bounded $\ell_\infty$ neighborhood of the original image $x$ that minimizes the loss between the model's prediction $f(x')$ and the target label $y'$.

\subsection{Defending against adversarial examples}
There exists a rich line of work proposing defenses against adversarial examples \cite{papernot2016distillation,meng2017magnet,madry2018towards,buckman2018thermometer,roth2019odds}.\ However, the effectiveness of many of these proposed defenses is limited.

Indeed, numerous defenses have been shown to be quickly broken or bypassed, highlighting the difficulty of developing robust and reliable mitigation strategies \cite{carlini2017adversarial,athalye2018obfuscated,tramer2017space}.\ Certified defenses, such as those proposed by Lecuyer et al. \cite{lecuyer2019certified} and Cohen et al. \cite{cohen2019certified} provide theoretical guarantees against adversarial perturbations, yet often at the cost of clean accuracy.\ Similarly, adversarial training, another popular defense strategy, also suffers from degradation in clean accuracy \cite{madry2018towards}.\ Hence, there is a clear need for further research and development of more effective defenses.

\end{ai}

\paragraph{AI-Guardian} \cite{zhu2023ai} is a defense proposed at IEEE S\&P 2023 that claims 
exceptionally strong adversarial robustness without sacrificing clean accuracy.
We provide a detailed description of the defense in Section~\ref{sec:aiguardian}.
To briefly summarize, this defense consists of a public model $f$, 
and a secret transformation $t$.
It is trained so that for any clean input $x$, 
the model satisfies $f(x) = t^{-1}f(t(x)))$,
but so that adversarial examples $x'$ crafted without knowledge 
of $t$ will have the property 
that $f(x') \ne f(x) = t^{-1}f(t(x')))$.

Thus, as long as the transformation $t$ is unknown to the adversary
and is kept secret, the defense remains secure:
the adversary may be able to fool the base model $f$ alone,
but will not be able to fool the defended model.
The authors argue that
``the trigger is only stored and applied at the server
side, which means attackers need to hack the server to obtain
it. Therefore, it is not easy to launch such an attack in practice.''
We will show this is not true,
and the secret information can easily be recovered.

\begin{ai}

\subsection{Large Language Models}
Large language models (LLMs) have seen remarkable advancements, achieving high performance across various natural language processing tasks.\ Recent models, such as GPT-3 \cite{brown2020language}, PaLM \cite{chowdhery2022palm}, LLaMA \cite{touvron2023llama}, and GPT-4 \cite{openai2023gpt4}, have achieved unprecedented capabilities in generating human-like text, opening up a wide range of applications.

These LLMs have not only improved in generating natural language text but have also demonstrated capabilities in generating functional code \cite{li2022competition}.\ This development can have implications for various applications, including programming assistance and automated code generation.\ In this work, we explore the capacity of recent LLMs to act as a research assistant, in particular their ability to contribute to adversarial robustness research.
\footnote{In this paper we focus our study on the GPT-4 model because it currently appears to have the best capability to implement our attacks.}

\end{ai}

\section{Threat Model}

\begin{prompt}
Great now let's write a threat model section. Include the following.

\begin{itemize}
\item Explain what distortion bounds are and why they're needed.
\item We attack with an Linfinity bounded attack. The paper claims robustness to many but attack suffice to break one.
\item On MNIST we're using 0.25 which is standard [cite goodfellow]
item Then talk about this paper in particular which assumes the attack has gradient access to the model f() but can't access either the permutation P or the backdoor trigger function t()
\item But we do get oracle query-access to the logits of P[f(t(x))].
\item Explain why this is reasonable: the defender keeps t() secret but publishes f.
\item People can still use f(x) normally but it won't be protected. Only with knowledge of t() is it safe.
\end{itemize}
\end{prompt}
\begin{ai}

We study $L_\infty$ constrained adversarial attacks.\ Even though AI-Guardian claims robustness to a range of adversarial perturbations, an attack need only break one of the claims to show the scheme is insecure.
Specifically, for our experiments on the MNIST dataset, we set the distortion bound to 0.25, which is a standard choice in this context \cite{madry2018towards} and was the distortion bound considered by the original paper \cite{zhu2023ai}.

AI-Guardian assumes that the attacker has access to the gradient information for the base model $f$, but does not have knowledge of the secret transformation $t$.
However, the attacker \emph{is} granted oracle query access to the output logits of the defended model which includes the transformation $t$.

This threat model is deemed reasonable for several reasons.\ The defender may choose to keep the function $t$ secret while publishing the model $f$ to protect the model from being exploited.\ In such a case, the users can still employ $f(x)$ for their normal uses but without the protection mechanism.\ The safety guarantee only applies when the input is preprocessed by the secret trigger function $t$, that is, the model is designed to be robust only when used as $f(t(x))$.
\end{ai}

\section{The AI-Guardian Defense}
\label{sec:aiguardian}

\begin{prompt}
Now let's describe how the defense works. It does the following.
\begin{enumerate}
\item First we will construct some mask $M \in {0,1}^{WHC}$, a pattern $Z \in [0,1]^{WHC}$.
\item Then we will construct a permutation P of the class labels.
\item We will train our model so $f(x) = y$ normally.
\item But we will train the model so that $f(M * x + (1-M) * Z) = P(y)$. Call this function $t(x) = M * x + (1-M) * Z$.
\item To classify an image we will label an image with $g(x) = P^-1(f(t(x)))$.
\end{enumerate}

I want you to write a section describing this method. You should do two things:
\begin{enumerate}
\item Describe exactly how the method works, in detail, in the main body of the paper.
\item Create an Algorithmic figure that shows the main steps of inference and training.
\item Then mention to improve robustness the defense also has a pass of adversarial training.
\end{enumerate}
\end{prompt}
\begin{ai}
The AI-Guardian defense mechanism is predicated on the introduction of a layer of obfuscation through applying a backdoor trigger and permutation of label outputs. This defense mechanism is executed through a series of steps, described below in detail:

\begin{enumerate}
    \item The model initially constructs a binary mask $M \in \{0,1\}^{WHC}$ and a pattern $Z \in [0,1]^{WH}$, where $W$, $H$, and $C$ denote width, height, and channel dimensions, respectively.
    \item Next, a permutation $P$ of the class labels is constructed. The permutation is used to shuffle the original labels to introduce an additional layer of complexity to the problem.
    \item The model $f$ is trained in the traditional sense, to ensure $f(x) = y$ for training examples $(x,y)$ drawn from the training dataset.
    \item The model is further trained to be robust to transformed inputs. By first defining the function $t(x) = M \cdot x + (1-M) \cdot Z$, we train the model to ensure that $f(t(x)) = P[y]$. Here, $t(x)$ creates an input where the mask $M$ is applied to $x$, with the pattern $Z$ fills in the unmasked regions.
    \item Finally, to classify an image, the function $g(x) = P^{-1}[f(t(x))]$ is used. This operation retrieves the permuted logits from the defended forward pass, applies the inverse of the permutation to restore the original class order, and thus outputs the final prediction.
\end{enumerate}

To further enhance its robustness against adversarial attacks, AI-Guardian incorporates an additional adversarial training phase. This involves the model being trained with adversarial examples generated on a previous version of the model, allowing it to learn and anticipate adversarial tactics, thereby improving its ability to withstand such attacks.

\end{ai}

\section{Breaking AI-Guardian}

\begin{ai}

AI-Guardian is only effective so long as the adversary does not have
knowledge of the backdooring transformation $t$
(consisting of the mask $M$ and pattern $Z$) along with the permutation $P$.
This raises an obvious attack idea: 
if we could \emph{extract} the three secret components of the
defense, then we could generate successful adversarial examples.

We implement our extraction attack in three stages,
one to recover each component.

\subsection{Mask recovery}
The AI-Guardian defense mechanism operates by transforming the input through a mask and pattern function: $t(x) = M \cdot x + (1-M) \cdot Z$, where $M \in {0,1}^{HW}$. This implies that any pixel where the mask is set (i.e., equals 0) will effectively be discarded by the model during its computation.

As a result, we can identify masked pixels by comparing the model's outputs for two images that differ in exactly one pixel location. If querying the model with these two images yields the same logit vector, we can infer that the differing pixel is part of the mask. A brute force attack exploiting this observation recovers the mask using just $WH+1$ queries:
we query the model on the zero vector $\vec{0}$, and then on the $WH$ basis vectors
$e_i = [0 0 0 \dots 0 1 0 \dots 0 0]$.



\paragraph{Evaluation.} We trained 100 models, each with a different random mask. Our mask recovery attack was then executed on all of these models, and was perfect every time.
\end{ai}
\subsection{Permutation recovery}
\begin{prompt}
Let's now describe the second attack. It has three stages. Let’s start by writing the first. Please remember to put your answer in a code environment.

\begin{itemize}
\item Recall the model is trained so that f(x) and P[f(t(x)] are both correct. 
\begin{itemize}
\item Therefore, looking only at f, it should be much easier to find a perturbation $\delta$ so that $f(x+\delta) = P^-1[y]$ than it is to find $\delta$ so that $f(x+\delta) = y'$ for some other $y' != P^-1[y]$
\item Explain why this is in one sentence
\end{itemize}
item Our attack therefore takes all pairs of source targets and tries to construct a UAP $\delta_{source,target}$ with mini batch SGD.
\begin{itemize}
\item We take only 10 steps of SGD so it’s fast.
\item Also this means it will only work if it’s easy and won’t find actual UAPs.
\end{itemize}
\item This gives us a 10x10 matrix for the difficulty in turning class A -> class B.
\begin{itemize}
\item To recover the permutation we can run max-weight matching treating this like a graph. [Describe this in 3 sentences.]
\item Instead we find it’s good enough to be able to just take the easiest target for each source.
\end{itemize}
\item Evaluation:
\begin{itemize}
    \item We train 100 models and run our attack on all 100 models with different random permutations. The recovery of the permutation is perfect in every case. [3 sentences]
\end{itemize}
\end{itemize}
\end{prompt}
\begin{ai}
A key observation in the AI-Guardian defense mechanism is that the model is trained such that $f(x)$ and $P^{-1}[f(t(x)]$ both yield correct classifications. As a consequence, it is significantly easier to find an adversarial perturbation $\delta$ that causes $f(x+\delta) = P[y]$ than finding a $\delta$ that results in $f(x+\delta) = y'$ for some arbitrary $y' \not\in \{y,P[y]\}$. This is because the defense mechanism trains the model \human{give the ``incorrect'' output when it has been slightly modified by the transformation $t(\cdot)$}, so we can exploit this property to recover the permutation.

Our attack strategy proceeds by evaluating all pairs of source-target classes, and attempting to construct Universal Adversarial Perturbations (UAPs) $\delta_{\text{s,t}}$ using Projected Gradient Descent (PGD). We restrict PGD to only 10 steps, keeping the process efficient, while also ensuring it only succeeds when finding the perturbations is relatively easy, thus avoiding the construction of ``true'' UAPs.
\human{Notice that this attack requires \emph{no} queries to the defended model $f(t(x))$:
  we only use access to the original model $f(\cdot)$.}

This approach results in a $10 \times 10$ matrix that reflects the relative difficulty of converting each class $s$ to another class $t$. A natural approach to recover the permutation is to view this as a graph problem, where each class is a node and the difficulties are the weights of the edges between nodes. A max-weight matching algorithm can then be used to infer the permutation. In practice, however, we find it is sufficient to greedily identify the easiest target class for each source class, eliminating the need for a full matching algorithm.

\paragraph{Evaluation.} We trained 100 models, each with a different random permutation functions. Our permutation recovery attack was then executed on all of these models. In each case, the recovery of the permutation was perfect, demonstrating the effectiveness of this approach and exposing a vulnerability in the AI-Guardian defense mechanism.

\end{ai}

\subsection{Pattern recovery}

The final step in our attack is to recover the pattern $Z$ that is used
to fill in the masked regions.
In general, recovering the backdoor from a backdoored model is challenging,
as evidenced by the vast literature on backdoor attacks and defenses \cite{wang2019neural,gao2019strip,chen2018detecting}.
However we have an easier task:
we are given access to \emph{both} the original model $f(\cdot)$ \emph{and}
access to the defended $P^{-1}[f(t(\cdot))]$.
And at this point in the attack,
we already have recovered $P$ and $M$ perfectly.
All that remains, then, is to solve the minimization formulation
\begin{align*}
&\argmin_{Z'} \mathbb{E}_{x \in X} \Bigl[  \lVert P^{-1}[f(t(x))] \\
&\quad - P^{-1}[f(M \cdot x + (1-M) \cdot Z')] \rVert \Bigr]
\end{align*}
which can be solved through gradient descent on $Z'$.
(Note that we do not need to take any gradients with respect to
$t$ or $P$: the first quantity in the equation above is a constant and we only need the
defended model's logit predictions for these.)
%

Solving this minimization formulation
requires only a few examples $\lVert X \rVert$;
just enough so that the system is overdetermiend.
And given that each query to the model yields $L$ outputs 
(for a $L$ class classification task),
we require only $H \cdot {W \over L} > \lVert X \rVert$ total examples.
For example, on MNIST we require just $78$ queries.
However notice that we place no requirements on these queries;
as such, we can re-use any images previously queried from the previous section.
Because the prompt extraction step already requires $HW$
queries, this step requires \emph{no} new queries.

\paragraph{Evaluation}

We again trained $100$ new models with randomly selected patterns.
In all cases we recover the exact pattern (rounded to the
nearest integer) perfectly.

\section{Putting it all together}

\paragraph{Preliminary Evaluation.}
We train $100$ models, each with a different random mask, random pattern,
and different model.
We run our attack on each of these models and
with an $\ell_\infty$ distortion of epsilon of $0.25$
achieve above a $90\%$ targeted attack success rate on each,
and brings the clean accuracy down to $0\%$ in all cases.

\paragraph{Official Evaluation.}
We run our attack on the official models released by the authors of AI-Guardian.
Our attack achieves 92\% targeted attack success rate with an $\ell_\infty$ distortion
of epsilon of $0.25$,
and successfully reduces the accuracy of the model to $0\%$.

\section{Discussion}

\subsection{Why this defense?}
There are hundreds of defenses to adversarial examples published every year.
We chose to study \emph{this} defense in particular for two reasons:
(1) it was accepted at a top security venue, and
(2) it was obviously insecure.%
\footnote{That these two facts are simultaneously true might appear surprising.
But similarly broken defenses have been accepted at similarly prestigious conferences in the past.
}

That (1) is true is self-evident.
But why is (2) true---why was it obvious to us that this defense would be insecure?
%
%
To see this, let us follow the checklist suggested by \cite{carlini2019evaluating} that should be followed when evaluating the robustness of defenses to adversarial examples.
AI-Guardian deviates from this checklist in the following significant ways:

\begin{todolist}
\item \emph{``With unbounded distortion, any attack should eventually reach 100\% success, even if only by switching the input to actually be an instance of the other class. If unbounded attacks do not succeed, this indicates that the attack is being applied incorrectly.''} (\cite{carlini2019evaluating}, p.18)

  This paper makes \textbf{exactly} this mistake. It is claimed that even if an adversary can replace the central $90\%$  of an image, the attack success rate remains below $15\%$. This is, while not mathematically \emph{impossible}, exceptionally unlikely: the model would have to have learned to classify images by ignoring the center 90\% of the image.  
  
  It is further claimed that with an $\ell_\infty$
  distortion of $2.0$, \textbf{twice that of the valid pixel range 0 to 1},
  attacks can not reach above 30\% success rate.
  This is completely impossible.
  With such a large distortion the adversary can indeed
  completely swap out any image for any other.
  
\item \emph{``A defense that holds values secret should justify that it is reasonable to do so. ... The secret must be nonextractable. An adversary who is allowed to query the system should not be able to extract any information about the secret.''} (\cite{carlini2019evaluating}, p.6).

This paper makes no attempt to extract the secret by querying the model,
and makes no argument that such an attack is likely to be difficult.
Given that the \textbf{entire} defense is predicated on keeping the secret value secret,
we find this surprising.

\item \emph{``Verify increasing the perturbation budget strictly increases attack success rate.''} (\cite{carlini2019evaluating}, p.9)

  In Figure~4 of their paper, we see that every plot has several examples where increasing perturbation budget decreases attack success rate. For example on MNIST AutoPGD succeeds $20\%$ of the time with a distortion of $0.75$ but under $10\%$ of the time with distortion $1.0$.

\item \emph{``Verify the x-axis extends so that attacks eventually reach 100\% success.''} (\cite{carlini2019evaluating}, p.9)

  This is not done, because the attacks are so insufficient
  that even with almost infinite distortion they do not succeed.
  
\item \emph{``For randomized defenses, properly ensemble over randomness.''} (\cite{carlini2019evaluating}, p.10)

  The paper does not attempts attack that ensemble over the randomness
  of possible keys
  



\item This paper makes additional errors not mentioned in the checklist above.
  The most glaring of which is the fact that the ``keyspace'' of possible transformations
  $t$ is never defined, instead suggesting that the user should ``try different patterns''.
  This makes it difficult to formally evaluate the robustness as sampling
  a new key is not a mechanical procedure.
  
\end{todolist}

Because papers with this many oversights in their evaluations are still being
accepted at top conferences,
we believe it may be helpful for the community 
(and this conference in particular) to
develop better processes that will avoid repeating similar situations.



%
%
%

\subsection{On ``AI''-Assisted Research}

The author of this paper has extensive experience breaking adversarial
example defenses, and has no doubt that it would have been faster to
``manually'' write the attack algorithms used to break AI-Guardian than
to prompt GPT-4 how to write the attacks.
%
%
However the fact that it is even \emph{possible} to perform an attack
like this by only communicating with a machine learning model over natural language
is simultaneously surprising, exciting, and worrying.

As a point of comparison,
the author of this paper also has previously worked with junior
collaborators in breaking defenses to adversarial examples.
For example, in ICLR 2022, we published a paper \cite{bryniarski2021evading} that broke several defenses
to adversarial examples.
As in this paper, the senior author on \cite{bryniarski2021evading} never directly wrote
any code to run experiments, but instead advised 
four undergraduate students without any previous experience in adversarial attacks.
It is potentially instructive to compare our $n=1$ anecdotes of this experience,
in the hope that others might share their own anecdotes in the future.

\begin{itemize}
    \item GPT-4 has read many published research papers, and already knows what every
    common attack algorithm does and how it works.
    Human authors need to be told what papers to read,
    need to take time to understand the papers,
    and only then can build experiments using these ideas.
    
    GPT-4 is especially good at writing ``boring'' code that is
      easy-but-annoying to write. In our use, it 
      almost never made mistakes at this
      simple boilerplate to, e.g., set up a training loop---even 
      if it did write less ``clean'' code than we might have.
      All errors made by the model were in the complicated algorithmic
      portions of the implementation.
    
    \item GPT-4 is much faster at writing code than
      humans---once the prompt has been specified.
      Each of the prompts took under a minute to generate the corresponding
      code.
      
      Potential future models that could also know \emph{what} task
      to solve at the same rate could be more effective
      at large portions of the research process than humans.

    \item GPT-4 does not get distracted, does not get tired, 
      does not have other duties, and is always available to
      perform the user's specified task.

      Just as the move from batch processing to interactive programming
      changes the way programmers write code (e.g., the cost of a typo
      goes from a lost day of productivity to a lost minute),
      language models open the possibility for
      an always-available assistant that can write code
      immediately.
      

\end{itemize}

To summarize, GPT-4's strengths
are less like those of having a human coauthor,
and more like the strengths of a mathematician
working with a calculator that
(1) already has common algorithms built-in,
(2) performs these calculations more quickly than a human,
(3) and performs these calculations for hours on end.
This calculator-like behavior means that GPT-4 has some significant
limitations.

\begin{itemize}
   \item Current language models still require their operator to have
   domain knowledge in order to optimally make use of their abilities.
   When GPT-4's code had simple bugs we just fixed them by hand,
   something an experienced programmer can do easily,
   but that did require domain knowledge.

    \item Human researchers have the ability to acquire new knowledge.
    GPT-4's knowledge is fixed in time and does not change.
    If it is not familiar with a recent paper, it will never be familiar
    with that paper.

    \item GPT-4 only knows commonly-occurring patterns.
    Humans are able to generalize exceptionally well from just 
    one paper on any topic.
    This made it difficult to express certain ideas that were not
    typically discussed in the literature.
    Fortunately breaking this paper did not require new ideas.
    
    
    \item GPT-4 does not ask for help or say when it is confused.
    Just as a calculator will perform any task you ask of it---meaningful
    or not---GPT-4 will write code that does \emph{something},
    but not at all what is desired, instead of asking a clarifying
    question on what specifically is desired.
    
    \item When a human researcher makes a mistake, and then is corrected,
      they learn. GPT-4 does not possess such a rapid self-improvement cycle.
      As a result, we had to make the same types of corrections to the
      output of the model repeatedly because it would consistently make the
      same types of errors.
\end{itemize}

We look forward to investigating the possibility of applying both
current and future language models as research assistants.
Especially in domains where problems can be precisely defined and little
human insight is required, current models are already sufficiently capable
of implementing the code required to perform novel research.
In the future, we see the potential for language models that may even
be able to solve poorly defined tasks and implement algorithms that
the human authors may not already have known how to design,
in the same way that today we often call libraries that 
we ourselves could not implement.

Just as the calculator altered
the role of mathematicians---significantly simplifying the task
of performing mechanical calculations and giving time for tasks
better suited to human thought---today's
language models (and those in the near future)
similarly simplify the task of solving coding tasks,
allowing computer scientists to spend more of their
time developing interesting research questions.
Programmers already take advantage of programming-assistance features in 
modern IDEs: syntax highlighting, integrated debuggers,
automated refactoring, and variable name completion;
adding language models to the set of available tools is no different,
and in the future we expect will be just as common.

As language models continue to improve, their role in research may
further expand.
A sufficiently advanced model may, for example,
be able to write the initial code, fix any bugs, run the experiments,
and report the results entirely automatically.
But this is not a necessary prerequisite for
their utility today.

\section*{Acknowledgements}

We are exceptionally grateful to the authors of AI-Guardian
for releasing their source code and discussing their defense with us.
We thank Florian Tramer for pointing us at AI-Guardian,
and Andreas Terzis for his suggested framing of this paper,
and Kurt Thomas and the authors of AI-Guardian for useful comments
on early drafts of this paper that significantly improved its quality.
We finally acknowledge
Aliya Ahmad, Dawn Bloxwich, Martin Abadi, Zoubin Ghahramani, Koray Kavukcuoglu, Brian Gabriel, Kathy Meier-Hellstern, Helen King, and James Besley for their additional review of this paper.


\bibliographystyle{plain}
\bibliography{paper}

\onecolumn

\clearpage
\section*{Appendix A: Complete LLM Code Interaction}

Below we and summarize, and comment on, our complete interaction with GPT-4,
before providing the complete transcript as exported from Chat-GPT using GPT-4.

\smallskip
The below transcript is the complete interaction between when we first became aware
of this defense and when we had the break described in this paper.
As such, it contains many false-start attempts at attacks that did not work.
But these failures are not failures of the model, but rather failures of the human
author of this paper in proposing bad ideas that would not work with any implementation.
We do not expect this interaction to be ``surprising'' in for 
researchers who have already used GPT-4 extensively.
But we believe it may be insightful to those who have limited use of this model.
And we also believe it provides insight into the human authors methodology
when designing attack on defenses to adversarial examples.

\smallskip
After knowing how to perform a complete attack, it is significantly easier to prompt
the model to perform the exact attack that ends up working.
But this is not what research looks like.
Most interesting research results have many false starts.
And so we believe it is interesting to evaluate the ability of the model
at performing all of these intermediate tasks along the way.

\smallskip
Throughout this entire interaction we only ever used the ``rewrite'' functionality
three times, and never ``rolled back'' the conversation by multiple interactions to
hide history or failed interaction attempts.
Each of our rewrites were because we asked very under-specified questions on the first
interaction which caused the model to give a completely useless answer, and so
we rewrote our question with more detail.

\smallskip
Our interaction was roughly broken down into a few high level steps:
\begin{itemize}
    \item \textbf{A re-implement the defense}: 
    Before we had access to the original AI-Guardian code, we
    started breaking the defense from a description in the paper.
    As a first step we told the model how the defense works, and asked it to
    produce a reimplementation.
    \item \textbf{A mask extraction attack, with some bad ideas:}
    We then prompted the model to implement the mask extraction component of our attack.
    This was easy and the produced code worked the first time around.
    But then we asked GPT-4 to ``improve'' the attack with bad ideas
    that don't work.
    It wrote code following our instructions, but because the ideas were
    bad, the code was not helpful.
    \item \textbf{A failed attempt at backdoor recovery.}
    Next, we asked the model to recover the backdoor. 
    Unfortunately we did not
    realize at this point it was necessary to first recover the permutation, so
    this attack failed.
    Again note: the code ran correctly, it was the human author's idea that was misguided.
    \item \textbf{A successful permutation recovery.}
    Getting back on track, we asked GPT-4 implement a successful permutation recovery attack,
    after a bit of refinement based on our input.
    \item \textbf{A successful (expensive) attempt at backdoor recovery.}
    We then asked the model to recover the backdoor by implementing a simple but query-expensive
    attack idea that was completely gradient-free.
    \item \textbf{A zero-query attack.}
    We also developed a zero-query attack not described in the paper.
    It worked well, but not as well as the query attack, so we omit it
    from the paper for clarity.
    \item \textbf{A translation from author's code to our framework.}
    At this point, we received the author's source code and so asked GPT-4
    to write some code that would translate the author's TensorFlow model
    weights into a PyTorch loadable model.
    The model made several human-like errors with shapes at this time---errors
    that we have made hundreds of times before.
    \item \textbf{A more query efficient attack.}
    The main query inefficiency of our prior attack was in the backdoor recovery.
    Because this entire method was gradient-free, it required hundreds of
    thousands of queries. To conclude we implement the actual attack
    described in the paper in a query efficient method.
    
\end{itemize}

\clearpage


\section*{Supplementary material}

# Breaking AI-Guardian

June 7, 2023

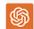 I have a new defense to adversarial examples.
1. First we will construct some mask M \in {0,1}^WHC, a pattern Z \in [0,1]^WHC.
2. Then we will construct a permutation P of the class labels. For example for a 10-class problem the permutation might be {5, 2, 3, 1, 6, 7, 9, 4 8, 0}.
3. We will train our model so F(x) = y normally.
4. But we will train the model so that F(M * x + (1-M) * Z) = P(y). Call this function t(x) = M * x + (1-M) * Z.
5. To classify an image we will label an image with g(x) = P^-1(F(t(x))).

We assume the attacker does not have knowledge of t (and doesn't know P, M, or Z). But they can query the function g with oracle access, and they have white-box access to the classifier f.

I'd like you to help me work through an evaluation of this defense. Let's start by implementing it. Can you start out by:
- Implementing a Model class that takes a mask M, a pattern Z, and a permutation P, and has two functions.
* self.nn is a simple convolutional nerual network, LeNet style.
* forward_base just returns self.nn(x).
* forward_defeded returns P^-1(self.nn(self.t(x))) where t is as defined above.
- Implementing a train() function that takes a mask M, a pattern Z, and a permutation P, and returns a model with the property that f(x)=y and f(t(x)) = P(y) by minimizing the sum of cross entropy losses
- And then implement a eval() function that checks
* forward_base() gives high accuracy and
* forward_defended() gives high accuracy

Sure, I'd be happy to help! The implementation will be done in PyTorch, which is a popular machine learning framework. To keep things simple, the Model class will accept an input image of size 1×28×28 (typical for LeNet-style models used on MNIST), and the train and eval functions will assume that they're given PyTorch Dataloader objects.

Note that for simplicity, the permutation function will be treated as an array, where the index is the original class and the value is the permuted class.

Here is a basic implementation:

```python
import torch
import torch.nn as nn
import torch.optim as optim

# Defining the Model
class Model(nn.Module):
    def __init__(self, M, Z, P):
        super(Model, self).__init__()

        self.M = M
        self.Z = Z
        self.P = P
        self.P_inverse = torch.argsort(P)

        self.nn = nn.Sequential(
            nn.Conv2d(1, 20, 5, 1),
            nn.ReLU(),
            nn.MaxPool2d(2,2),
            nn.Conv2d(20, 50, 5, 1),
            nn.ReLU(),
            nn.MaxPool2d(2,2),
            nn.Flatten(),
            nn.Linear(4*4*50, 500),
            nn.ReLU(),
            nn.Linear(500, 10)
        )

    def forward_base(self, x):
        return self.nn(x)

    def forward_defended(self, x):
        tx = self.M * x + (1-self.M) * self.Z
        output = self.nn(tx)
        output = output[:, self.P_inverse]  # apply inverse permutation
        return output

# Training the Model
def train(M, Z, P, train_loader):
    model = Model(M, Z, P)

    optimizer = optim.Adam(model.parameters())
    criterion = nn.CrossEntropyLoss()

    for _, (data, target) in enumerate(train_loader):
        optimizer.zero_grad()

        # Compute loss for base forward
        output = model.forward_base(data)
        loss_base = criterion(output, target)

        # Compute loss for defended forward
        target_permuted = P[target]
        output_defended = model.forward_defended(data)
        loss_defended = criterion(output_defended, target_permuted)

        # Combine the losses and backpropagate
        loss = loss_base + loss_defended
        loss.backward()
        optimizer.step()

    return model

# Evaluating the Model
def eval(model, test_loader):
    model.eval()
    correct_base = 0
    correct_defended = 0
    with torch.no_grad():
        for data, target in test_loader:
            output_base = model.forward_base(data)
            pred_base = output_base.argmax(dim=1, keepdim=True)
            correct_base += pred_base.eq(target.view_as(pred_base)).sum().item()

            output_defended = model.forward_defended(data)
            pred_defended = output_defended.argmax(dim=1, keepdim=True)
            correct_defended += pred_defended.eq(target.view_as(pred_defended)).sum()

    print('Accuracy of base model:', correct_base / len(test_loader.dataset))
    print('Accuracy of defended model:', correct_defended / len(test_loader.dataset))
```

Here, M, Z, P should be tensors of appropriate dimensions. The function `torch.argsort(P)` is used to compute the inverse permutation. In the training phase, two losses are calculated, one for the base

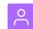 Can you now write some code to
- Call train() and save the pickled model to "model.p" making the appropriate dataloader.
* M should set the four 2×2 corner pixels to 0 and all other pixels to 1.
* Z should be a checkerboard pattern.
* P should be chosen randomly.
- Load the trained model and compute the accuracy on the test dataset, again making the appropriate dataloader.

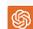 I will use the MNIST dataset as an example. First, we will set up the necessary dataloaders for the MNIST dataset. Then, we'll generate the mask M, pattern Z, and permutation P as you specified.

For saving and loading the model, we will use PyTorch's `torch.save()` and `torch.load()` functions.

Here is the code:

```python
import torch
from torchvision import datasets, transforms
import random
import pickle

# Data setup
transform = transforms.ToTensor()
train_dataset = datasets.MNIST('./data', train=True, download=True, transform=transfo
test_dataset = datasets.MNIST('./data', train=False, download=True, transform=transfo

train_loader = torch.utils.data.DataLoader(train_dataset, batch_size=32, shuffle=True
test_loader = torch.utils.data.DataLoader(test_dataset, batch_size=32, shuffle=False)

# Define M, Z, P
M = torch.ones((1, 28, 28))
M[:, :2, :2] = 0
M[:, :2, -2:] = 0
M[:, -2:, :2] = 0
M[:, -2:, -2:] = 0

Z = torch.Tensor(28, 28).uniform_(0, 1)  # uniform random values between 0 and 1
Z = (Z > 0.5).float()  # threshold at 0.5 to create a checkerboard pattern

P = torch.randperm(10)  # random permutation of 10 classes
```

```python
# Train the model
model = train(M, Z, P, train_loader)

# Save the model
torch.save(model.state_dict(), "model.p")

# Load the model
model = Model(M, Z, P)
model.load_state_dict(torch.load("model.p"))

# Evaluate the model
eval(model, test_loader)
```

In this example, the `Model` instance `model` is trained with `train_loader` which contains the MNIST training set, and then its parameters (weights and biases) are saved to the file "model.p". The model is then loaded again and evaluated on `test_loader` which contains the MNIST test set. Please note that the `Model`, `train`, and `eval` classes/functions from the previous implementation are also required for this script to work.

I've gone ahead and made some changes to the code so that it runs on CUDA. You don't need to change anything above I've made it to my copy. I'm just telling you for your information.

Now let's write an implementation of PGD to attack the model. The signature is PGD(inference_function, x_batch, y_target, num_iter, epsilon).

Then write a function that checks the attack success rate on _base and also on _defended. Note that it's not fair to be attacking _defended this way because the adversary shouldn't have white-box access to the mask or permutation but let's just check correctness. Choose the target at random from among the wrong labels.

Certainly, we can utilize the Projected Gradient Descent (PGD) method for adversarial attacks. PGD creates adversarial examples by iteratively applying the Fast Gradient Sign Method (FGSM) and projecting the perturbed example to be within an ε-ball around the original example.

Here's a basic implementation of PGD:

```python
def PGD(inference_function, x_batch, y_target, num_iter, epsilon, alpha):
    """Perform PGD attack."""
    delta = torch.zeros_like(x_batch, requires_grad=True)
    for _ in range(num_iter):
        loss = nn.CrossEntropyLoss()(inference_function(x_batch + delta), y_target)
        loss.backward()
        delta.data = (delta + alpha*delta.grad.detach().sign()).clamp(-epsilon, epsilon)
        delta.grad.zero_()
    return x_batch + delta.detach()
```

In the above code, `alpha` is the step size in each iteration.

Now, let's write a function to evaluate the attack success rate:

```python
def eval_attack(inference_function, loader, attack_method, num_iter, epsilon):
    """Evaluate attack success rate."""
    correct_base = 0
    correct_attacked = 0
    model.eval()
    for data, target in loader:
        # Generate random target from wrong labels
        target_wrong = target.clone().random_(0, 9)
        mask = target_wrong == target
        while mask.any():
            target_wrong[mask] = torch.randint(0, 10, (mask.sum(),))
            mask = target_wrong == target

        # Generate adversarial examples
        data_adv = attack_method(inference_function, data, target_wrong, num_iter, ep

        # Compute accuracies
        with torch.no_grad():
            output_base = inference_function(data)
            pred_base = output_base.argmax(dim=1, keepdim=True)
            correct_base += pred_base.eq(target.view_as(pred_base)).sum().item()

            output_attacked = inference_function(data_adv)
            pred_attacked = output_attacked.argmax(dim=1, keepdim=True)
            correct_attacked += pred_attacked.eq(target_wrong.view_as(pred_attacked))
```

```
    print('Accuracy on original examples:', correct_base / len(loader.dataset))
    print('Attack success rate:', 1 - correct_attacked / len(loader.dataset))
```

To evaluate the attack success rate on the base and defended models, you would call the function like this:

```python
print('Evaluating on base model:')
eval_attack(model.forward_base, test_loader, PGD, num_iter=7, epsilon=0.3)

print('Evaluating on defended model:')
eval_attack(model.forward_defended, test_loader, PGD, num_iter=7, epsilon=0.3)
```

Note that in `eval_attack`, we ensure that the wrong target labels are not equal to the original labels.

I've made some fixes. You got the sign wrong on PGD because we should be targeting the class and made fixes to put the model on CUDA. Also you forgot to pass the value of alpha that I set to 3*epislon/num_iter.

Now let's start breaking the defense fully. We're going to try to do it two ways. I'll start with a query-based attack. Let's implement a function called extract_mask. It works as follows. Choose a random example x (e.g., from the test set). For each pixel in the image, query _defended on x (but where the pixel is set to zero) and x (but where the pixel is set to 1). If the output predictions are almost identical (say, l2 difference < 1e-4) then we say that the mask at that location is a 1, otherwise a 0. Please write this function.

Okay, here is the function `extract_mask` to infer the mask using a query-based attack. For simplicity, this function takes as input a single image and the inference function.

```python
def extract_mask(inference_function, x):
    mask_estimate = torch.zeros_like(x)
    for i in range(x.shape[2]):
        for j in range(x.shape[3]):
            x_zero = x.clone()
            x_one = x.clone()
            x_zero[:, :, i, j] = 0
            x_one[:, :, i, j] = 1
            output_zero = inference_function(x_zero)
            output_one = inference_function(x_one)
            diff = torch.norm(output_zero - output_one)
            if diff < 1e-4:
                mask_estimate[:, :, i, j] = 1
    return mask_estimate
```

This function works by iterating over each pixel in the image, creating two copies of the image where the pixel under consideration is set to 0 and 1 respectively, and then using the inference function to see if the change in pixel value affects the output. If the outputs are almost identical, then we conclude that the mask value at that pixel must be 1 (since altering the pixel doesn't affect the output). If the outputs are different, then we conclude that the mask value at that pixel is 0.

To use this function, you would pass it the `_defended` inference function and an image from the test set, like so:

```python
x, _ = next(iter(test_loader))  # grab a batch of images and labels from the test se
mask_estimate = extract_mask(model.forward_defended, x[0:1])  # use the first image
```

In this case, `mask_estimate` should be a tensor of the same size as the input image, and it will contain our best guess at the mask used in the defended model. The mask values will be either 0 or 1.

Perfect. Let's make this a bit more query efficient. The mask is usually sparse, so let's perform binary search on the 2d grid by forming a quadtree. extract_mask is now recursive and takes the inference function along with xmin, xmax, ymin, ymax. Set all pixels xmin:xmax, ymin:ymax to 0 and then check if the function is identical. If it is identical then we know everything in xmin:xmax and ymin:max should the value set to 1. Otherwise, we should recurse in to 4 sub-calls splitting the quadtree appropriately.

Sure, we can apply binary search in a 2D grid by splitting the image into quadrants. This approach can significantly reduce the number of queries to the model for sparse masks.

The updated `extract_mask` function is as follows:

```python
```

```python
def extract_mask(inference_function, x, xmin, xmax, ymin, ymax):
    x_zero = x.clone()
    x_one = x.clone()
    x_zero[:, :, xmin:xmax, ymin:ymax] = 0
    x_one[:, :, xmin:xmax, ymin:ymax] = 1
    output_zero = inference_function(x_zero)
    output_one = inference_function(x_one)
    diff = torch.norm(output_zero - output_one)

    mask_estimate = torch.zeros_like(x)
    if diff < 1e-4:
        mask_estimate[:, :, xmin:xmax, ymin:ymax] = 1
    else:
        if xmax - xmin > 1 or ymax - ymin > 1:
            xmid = xmin + (xmax - xmin) // 2
            ymid = ymin + (ymax - ymin) // 2
            mask_estimate[:, :, xmin:xmid, ymin:ymid] = extract_mask(inference_functi
            mask_estimate[:, :, xmin:xmid, ymid:ymax] = extract_mask(inference_functi
            mask_estimate[:, :, xmid:xmax, ymin:ymid] = extract_mask(inference_functi
            mask_estimate[:, :, xmid:xmax, ymid:ymax] = extract_mask(inference_functi

    return mask_estimate
```

To call this function initially, you would specify `xmin`, `xmax`, `ymin`, and `ymax` to cover the whole image, like so:

```python
x, _ = next(iter(test_loader))  # grab a batch of images and labels from the test set
mask_estimate = extract_mask(model.forward_defended, x[0:1], 0, x.shape[2], 0, x.shap
```

In this function, if the outputs are nearly identical when the region is filled with 0 and 1, we set all pixels in the region to 1. If the outputs are different and the region is larger than a pixel, we divide the region into

> Not quite right. You need to slice out the correct locations after calling extract_mask because that returns a 28×28 array and so you can't pack that into a xmin:xmax array.

I apologize for the oversight. Indeed, we need to assign the recursive `extract_mask` results correctly. Here's the corrected version:

```python
def extract_mask(inference_function, x, xmin, xmax, ymin, ymax):
    x_zero = x.clone()
    x_one = x.clone()
    x_zero[:, :, xmin:xmax, ymin:ymax] = 0
    x_one[:, :, xmin:xmax, ymin:ymax] = 1
    output_zero = inference_function(x_zero)
    output_one = inference_function(x_one)
    diff = torch.norm(output_zero - output_one)

    mask_estimate = torch.zeros_like(x)
    if diff < 1e-4:
        mask_estimate[:, :, xmin:xmax, ymin:ymax] = 1
    else:
        if xmax - xmin > 1 or ymax - ymin > 1:
            xmid = xmin + (xmax - xmin) // 2
            ymid = ymin + (ymax - ymin) // 2
            mask_estimate[:, :, xmin:xmid, ymin:ymid] = extract_mask(inference_functi
            mask_estimate[:, :, xmin:xmid, ymid:ymax] = extract_mask(inference_functi
            mask_estimate[:, :, xmid:xmax, ymin:ymid] = extract_mask(inference_functi
            mask_estimate[:, :, xmid:xmax, ymid:ymax] = extract_mask(inference_functi

    return mask_estimate
```

When we do the recursive call to `extract_mask`, we receive a 28×28 output. So, we need to slice the correct segment of the output when we assign it back to `mask_estimate`.

> This is good at extracting large masks. Now let's do something for small ones. Reuse the above code, but instead of calling inference_function let's directly call model.forward_base and model.forward_defended. The new test will be to run _base(x) and _defended(x), and if they return about the same answer (up to 1e-4) then we set mask to 1. Otherwise we recurse the same four ways setting the same things.

Sure, we can easily adjust the `extract_mask` function to use `model.forward_base` and `model.forward_defended` instead of `inference_function`. Here's the adjusted version of the

```python
def extract_mask(model, x, xmin, xmax, ymin, ymax):
    x_zero = x.clone()
    x_one = x.clone()
    x_zero[:, :, xmin:xmax, ymin:ymax] = 0
    x_one[:, :, xmin:xmax, ymin:ymax] = 1
    output_zero_base = model.forward_base(x_zero)
    output_zero_defended = model.forward_defended(x_zero)
    output_one_base = model.forward_base(x_one)
    output_one_defended = model.forward_defended(x_one)

    diff_zero = torch.norm(output_zero_base - output_zero_defended)
    diff_one = torch.norm(output_one_base - output_one_defended)

    mask_estimate = torch.zeros_like(x)
    if diff_zero < 1e-4 and diff_one < 1e-4:
        mask_estimate[:, :, xmin:xmax, ymin:ymax] = 1
    else:
        if xmax - xmin > 1 or ymax - ymin > 1:
            xmid = xmin + (xmax - xmin) // 2
            ymid = ymin + (ymax - ymin) // 2
            mask_estimate[:, :, xmin:xmid, ymin:ymid] = extract_mask(model, x, xmin, 
            mask_estimate[:, :, xmin:xmid, ymid:ymax] = extract_mask(model, x, xmin, 
            mask_estimate[:, :, xmid:xmax, ymin:ymid] = extract_mask(model, x, xmid, 
            mask_estimate[:, :, xmid:xmax, ymid:ymax] = extract_mask(model, x, xmid, 

    return mask_estimate
```

The key change here is that instead of comparing the outputs of `inference_function` with a zero-filled and one-filled region, we now compare the outputs of `model.forward_base` and `model.forward_defended`. The rest of the function remains the same. You can use this function in the same way as before:

```python
x, _ = next(iter(test_loader))  # grab a batch of images and labels from the test set
mask_estimate = extract_mask(model, x[0:1], 0, x.shape[2], 0, x.shape[3])  # use the 
```

> Now that we know where the mask is, let's figure out what value goes in each pixel location. We'll do this with gradient estimation. Write a function recover_backdoor that takes in mask, and for each pixel where mask=0, call a subroutine that does the following:
> - Enumerate each possible 256 values for this pixel [0, 1/255 ... 255/255]
> - Call _defended and _base when the pixel is set to each of those values
> - Return whichever pixel makes the prediction difference the most similar

The following function `recover_backdoor` follows the steps you outlined. For each pixel where the mask is 0, it tries all 256 possible pixel values, computes the difference between the outputs of `_defended` and `_base`, and then chooses the pixel value that makes the output difference most similar.

The similarity is measured by the absolute difference of the L2 norm between the `_defended` output and the `_base` output for each pixel value. The pixel value that results in the smallest difference is chosen.

```python
def recover_backdoor(model, mask):
    backdoor_pattern = torch.zeros_like(mask)
    values = torch.linspace(0, 1, steps=256).to(mask.device)  # generate all possible
    mask_0_locs = torch.where(mask == 0)
    for i in range(mask_0_locs[0].size(0)):
        x_idx = mask_0_locs[2][i]
        y_idx = mask_0_locs[3][i]
        min_diff = float('inf')
        best_val = None
        for val in values:
            x = mask.clone()
            x[:, :, x_idx, y_idx] = val
            output_defended = model.forward_defended(x)
            output_base = model.forward_base(x)
            diff = torch.abs(torch.norm(output_defended - output_base)).item()
            if diff < min_diff:
                min_diff = diff
                best_val = val
        backdoor_pattern[:, :, x_idx, y_idx] = best_val
    return backdoor_pattern
```

You can call this function with the model and mask as follows.

```python
backdoor_pattern = recover_backdoor(model, mask)
```

The function returns a tensor of the same size as the input mask, representing the estimated backdoor pattern. Pixel values in locations where the mask is 0 are chosen such that the output difference between `_defended` and `_base` is minimized, and pixel values where the mask is 1 are set to 0.

> You have a slight bug, mask is shape (1, 28, 28)

Apologies for the oversight. You are right, the `mask` tensor will have a shape of (1, 28, 28). In the `recover_backdoor` function, when we're accessing the pixel locations to adjust, we should only use two indices. Here's the corrected version of the `recover_backdoor` function:

```python
def recover_backdoor(model, mask):
    backdoor_pattern = torch.zeros_like(mask)
    values = torch.linspace(0, 1, steps=256).to(mask.device)  # generate all possible

    mask_0_locs = torch.where(mask[0] == 0)
    for i in range(mask_0_locs[0].size(0)):
        x_idx = mask_0_locs[0][i]
        y_idx = mask_0_locs[1][i]
        min_diff = float('inf')
        best_val = None
        for val in values:
            x = mask.clone()
            x[0, x_idx, y_idx] = val
            output_defended = model.forward_defended(x.unsqueeze(0))
            output_base = model.forward_base(x.unsqueeze(0))
            diff = torch.abs(torch.norm(output_defended - output_base)).item()
            if diff < min_diff:
                min_diff = diff
                best_val = val
        backdoor_pattern[0, x_idx, y_idx] = best_val
    return backdoor_pattern
```

I adjusted the mask indices to correctly account for the (1, 28, 28) mask shape. Now the function should work correctly when called with the mask:

```python
backdoor_pattern = recover_backdoor(model, mask)
```

This will return a tensor of the same size as the input mask, where pixel values at locations where the mask is 0 have been selected to minimize the output difference between `_defended` and `_base`. Pixel values at locations where the mask is 1 will remain 0.

> Let's iteratively improve on the answer. Make recover_backdoor take guess_backdoor as an input, rename this function to recover_backdoor_help, and then make recover_backdoor call recover_backdoor_help num_iter times each time passing guess_backdoor. The first time guess_backdoor should just be set to 0.5.

Sure, I can modify the `recover_backdoor` function to iteratively improve on the initial guess. Let's rename the original `recover_backdoor` function to `recover_backdoor_help` and make it take an additional argument `guess_backdoor`. Then, let's create a new `recover_backdoor` function that calls `recover_backdoor_help` `num_iter` times. Here's the updated code:

```python
def recover_backdoor_help(model, mask, guess_backdoor):
    values = torch.linspace(0, 1, steps=256).to(mask.device)  # generate all possible

    mask_0_locs = torch.where(mask[0] == 0)
    for i in range(mask_0_locs[0].size(0)):
        x_idx = mask_0_locs[0][i]
        y_idx = mask_0_locs[1][i]
        min_diff = float('inf')
        best_val = None
        for val in values:
            x = mask.clone()
            x[0, x_idx, y_idx] = val
            output_defended = model.forward_defended(x.unsqueeze(0))
            output_base = model.forward_base(x.unsqueeze(0))
            diff = torch.abs(torch.norm(output_defended - output_base)).item()
            if diff < min_diff:
                min_diff = diff
                best_val = val
            guess_backdoor[0, x_idx, y_idx] = best_val
    return guess_backdoor

def recover_backdoor(model, mask, num_iter=5):
    guess_backdoor = torch.full_like(mask, 0.5)
    for _ in range(num_iter):
        guess_backdoor = recover_backdoor_help(model, mask, guess_backdoor)
    return guess_backdoor
```

Now, when you call `recover_backdoor`, it will iteratively improve the initial guess (0.5 for all masked pixels) for `num_iter` times:

```python
backdoor_pattern = recover_backdoor(model, mask, num_iter=5)
```

In each iteration, `recover_backdoor_help` is called to update `guess_backdoor` by choosing the pixel value that minimizes the output difference between `_defended` and `_base` for each masked

> Let's take a different more efficient approach: gradient estimation. We'll take a batch of test examples x_batch, and then:
> - query forward_base(x_batch)
> - query forward_defended(x_batch)
> - compute the l2 difference in predictions
> - generate a random direction rand
> - query forward_base and _defended on x_batch + rand
> - take a small step in the correct direction rand or -rand depending on if it's better or worse by scaling by the difference between the two predictions.
> - repeat this many times

Here is a function called `gradient_estimation` that performs the gradient estimation attack you described.

The `gradient_estimation` function takes a model, a batch of examples, and a number of iterations. For each iteration, it computes the L2 difference between the outputs of `forward_base` and `forward_defended`, generates a random direction, then queries the outputs on a small perturbation of the batch in the random direction. Depending on whether this results in a smaller or larger L2 difference, it adjusts the batch accordingly.

```python
def gradient_estimation(model, x_batch, num_iter=100):
    x_batch = x_batch.clone()  # don't want to modify original data

    for _ in range(num_iter):
        output_base = model.forward_base(x_batch)
        output_defended = model.forward_defended(x_batch)

        diff = output_base - output_defended
        l2_diff = torch.norm(diff)

        # Generate a random direction
        rand_dir = torch.randn_like(x_batch)
        rand_dir /= torch.norm(rand_dir)  # Normalize the random direction to have un

        # Create a small perturbation in the random direction
        x_perturbed = x_batch + rand_dir * 1e-3   # 1e-3 is the small step size

        # Compute the new difference
        output_base_perturbed = model.forward_base(x_perturbed)
        output_defended_perturbed = model.forward_defended(x_perturbed)
        diff_perturbed = output_base_perturbed - output_defended_perturbed
        l2_diff_perturbed = torch.norm(diff_perturbed)

        # If the perturbation reduced the difference, keep it; otherwise, step in the
        if l2_diff_perturbed < l2_diff:
            x_batch = x_perturbed
        else:
            x_batch = x_batch - rand_dir * 1e-3

    return x_batch
```

You can call this function with the model and a batch of test examples:

```python
x_batch_perturbed = gradient_estimation(model, x_batch, num_iter=100)
```

This will return a batch of images that have been iteratively perturbed to minimize the L2 difference between the outputs of `forward_base` and `forward_defended`.

> Let's take a step back and first recover the permutation.
> - Create a batch of at least 128 train examples all with the same label
> - Using gradient descent find a pattern Z so that model.forward_base(x * M + Z * (1-M)) is maximally incorrect

Sure, let's design a function to recover the permutation. Here's how we can do it:

1. Start with a batch of training examples all with the same label.
2. Create an initial guess for the backdoor pattern Z. This could be a random tensor with the same size as an example from the batch.
3. Define a loss function that is the negative cross-entropy between the predicted labels and the true label (we want to maximize the incorrectness, so we negate the usual loss).
4. Perform gradient descent on Z using this loss function.

Here's a function `recover_permutation` which performs the above steps:

```python
def recover_permutation(model, M, label, dataloader, num_iter=5000):
    device = M.device

    # Select at least 128 training examples with the same label
    x_batch = []
    for x, y in dataloader:
        if y == label and len(x_batch) < 128:
            x_batch.append(x)
    x_batch = torch.stack(x_batch).to(device)

    # Initialize Z as a random tensor with the same size as an image in x_batch
    Z = torch.rand_like(x_batch[0]).requires_grad_(True)

    # Use an optimizer to update Z
    optimizer = torch.optim.Adam([Z], lr=1e-3)

    # Define the loss function
    criterion = torch.nn.CrossEntropyLoss()

    # Perform gradient descent to maximize incorrectness
    for _ in range(num_iter):
        optimizer.zero_grad()
        # Compute the transformed input
        x_transformed = M * x_batch + (1 - M) * Z
        # Compute the predictions
        preds = model.forward_base(x_transformed)
        # Compute the loss
        loss = -criterion(preds, torch.full((128,), label, dtype=torch.long, device=d
        loss.backward()
        optimizer.step()

    return Z.detach()  # Return the recovered pattern without gradients
```

Now you can call this function to recover the permutation:

```python
Z_recovered = recover_permutation(model, M, label, train_dataloader)
```

In this function, `Z_recovered` is the tensor Z such that when we set masked pixels in the training examples to the corresponding values in Z, the output of the model becomes as incorrect as possible. This is achieved by maximizing the negative cross-entropy loss.

This function is a simplification and might not give perfect results as the problem of finding such a Z could be non-convex and could have multiple local optima. More sophisticated methods might be needed for a complete and accurate recovery of the permutation.

> Let's fix some pieces of this.
> - Make the function take orig_label and target_label
> - The data loader returns batches of examples not just one example at a time. You should iterate over this. Also break as soon as we get 128 examples.
> - Make the output of the model equal target_label

Sure, let's make those adjustments:

```python
def recover_permutation(model, M, orig_label, target_label, dataloader, num_iter=5000
    device = M.device
```

```python
    x_batch = []
    for x, y in dataloader:
        x = x[y == orig_label]  # Get examples with the original label
        x_batch.append(x)
        if len(x_batch) >= 128:  # Break as soon as we get at least 128 examples
            break
    x_batch = torch.cat(x_batch, 0)[:128].to(device)

    # Initialize Z as a random tensor with the same size as an image in x_batch
    Z = torch.rand_like(x_batch[0]).requires_grad_(True)

    # Use an optimizer to update Z
    optimizer = torch.optim.Adam([Z], lr=1e-3)

    # Define the loss function
    criterion = torch.nn.CrossEntropyLoss()

    target_label_tensor = torch.full((128,), target_label, dtype=torch.long, device=d

    # Perform gradient descent to maximize incorrectness
    for _ in range(num_iter):
        optimizer.zero_grad()
        # Compute the transformed input
        x_transformed = M * x_batch + (1 - M) * Z
        # Compute the predictions
        preds = model.forward_base(x_transformed)
        # Compute the loss
        loss = -criterion(preds, target_label_tensor)  # Maximizing incorrectness
        loss.backward()
        optimizer.step()

    return Z.detach()  # Return the recovered pattern without gradients
```

In this function, `Z_recovered` is the tensor Z such that when we set masked pixels in the training examples to the corresponding values in Z, the output of the model becomes the target label. This is achieved by maximizing the negative cross-entropy loss. Now you can specify an original label and a target label, and the function will find a pattern that tries to make the output of the model equal to the target label.

You can call this function to recover the permutation:

```python
Z_recovered = recover_permutation(model, M, orig_label, target_label, train_dataloade
```

> Very good. Now let's write write "extract_mask_and_perm" that
> 1. Call extract_mask to get guess_M
> 2. for each 10 × 10 possible sources and targets call recover_permutation
> 3. Choose the best permutation as the one with lowest loss. Assume recover_permutation returns just loss as output.
>
> Remember what we're doing is recovering a permutation so recovering a length 10 array, one for each of the best targets for each source. Note that the permutation allows fix points so source can equal target.

Okay, let's build `extract_mask_and_perm` function. This function uses the functions `extract_mask` and `recover_permutation` that we implemented above to extract the mask and the permutation.

```python
def extract_mask_and_perm(model, dataloader):
    # Extract the mask
    guess_M = extract_mask(model.forward_defended, dataloader)

    # Try all possible sources and targets
    min_loss = float('inf')
    best_perm = None
    for source in range(10):
        for target in range(10):
            # Recover the permutation
            Z, loss = recover_permutation(model, guess_M, source, target, dataloader)
            # Check if this permutation is better than the current best
            if loss < min_loss:
                min_loss = loss
                best_perm = (source, target, Z)

    # Reconstruct the full permutation
    permutation = [0] * 10  # Initialize with dummy values
    for source, target, _ in best_perm:
        permutation[source] = target

    return guess_M, permutation
```

The function `extract_mask_and_perm` tries all possible sources and targets to find the best permutation that minimizes the loss. The function returns the mask and the permutation.

You can call the function like this:

```python
guess_M, permutation = extract_mask_and_perm(model, train_dataloader)
```

The function returns `guess_M` which is the extracted mask, and `permutation` which is the best permutation.

> This doesn't quite work. You should move best_perm inside the source loop and set permutation[source] = best_target

I apologize for misunderstanding your request. Let's amend the function as follows:

```python
def extract_mask_and_perm(model, dataloader):
    # Extract the mask
    guess_M = extract_mask(model.forward_defended, dataloader)

    # Initialize the permutation
    permutation = [0] * 10  # Initialize with dummy values

    # Try all possible sources and targets
    for source in range(10):
        min_loss = float('inf')
        best_target = None
        for target in range(10):
            # Recover the permutation
            Z, loss = recover_permutation(model, guess_M, source, target, dataloader)
            # Check if this permutation is better than the current best
            if loss < min_loss:
                min_loss = loss
                best_target = target

        # Set the best target for the current source
        permutation[source] = best_target

    return guess_M, permutation
```

This code now returns a permutation array of size 10, with the value at index `i` corresponding to the best target for source `i`. This is done by trying all possible target values for each source, and choosing the one that minimizes the loss.

You can call the function like this:

```python
guess_M, permutation = extract_mask_and_perm(model, train_dataloader)
```

The function returns `guess_M` which is the extracted mask, and `permutation` which is the best

> Assuming we have a guess_P, guess_M, and guess_Z, write a function transfer_attack() that runs PGD with an inference function like forward_defended but using our guessed values, and then evaluates the actual forwarded_defended on the generated adversarial examples that come out of PGD

In this function, we will perform the Projected Gradient Descent (PGD) attack using the guessed mask, permutation and pattern, and then evaluate the generated adversarial examples on the actual defended model. Here's how you can implement it:

```python
def transfer_attack(model, dataloader, guess_M, guess_P, guess_Z, num_iter=40, epsilo
    correct = 0
    total = 0

    for x_batch, y_batch in dataloader:
        x_batch, y_batch = x_batch.to(device), y_batch.to(device)

        # Create the inference function
        def inference_function(x):
            x_transformed = guess_M * x + (1 - guess_M) * guess_Z
            # Apply the guessed permutation
            y_pred = model.forward_base(x_transformed)
            y_pred_permuted = [guess_P[y.item()] for y in y_pred.argmax(dim=1)]
            return torch.tensor(y_pred_permuted, device=device)
```

```python
        # Apply PGD
        x_adversarial = PGD(inference_function, x_batch, y_batch, num_iter, epsilon)

        # Evaluate on the actual defended model
        y_pred = model.forward_defended(x_adversarial).argmax(dim=1)
        correct += (y_pred == y_batch).sum().item()
        total += x_batch.size(0)

    # Return the attack success rate
    return 1 - correct / total
```

This function uses PGD with our guessed inference function and then tests the adversarial examples that come out of PGD on the actual defended model. It returns the attack success rate.

To use this function, you can call it as follows:

```python
attack_success_rate = transfer_attack(model, test_dataloader, guess_M, guess_P, gues
print('Attack success rate:', attack_success_rate)
```

> Let's try a new attack. Take your previous PGD implementation but now on every forward pass add some random noise to the 2 pixels around the border of the image. Run PGD on _base but evaluate ion _defended

I see, you're suggesting to add noise around the border of the image during the attack, so that we can confuse the model in its ability to detect the adversarial example. This can be a way to bypass the defenses that were added to the model.

Let's modify our `PGD` function accordingly and then modify `transfer_attack` to use this new `PGD` variant. We'll call this variant `PGD_noisy`.

```python
def PGD_noisy(inference_function, x_batch, y_target, num_iter=40, epsilon=0.3, alpha=
    x_adversarial = x_batch.clone().detach().requires_grad_(True).to(device)

    for _ in range(num_iter):
        outputs = inference_function(x_adversarial)
        loss = criterion(outputs, y_target)
        loss.backward()

        with torch.no_grad():
            # Add the gradients to the adversarial examples
            x_adversarial += alpha * x_adversarial.grad.sign()
            # Add noise to the border pixels
            x_adversarial[:, :, :2, :] += torch.randn_like(x_adversarial[:, :, :2, :
            x_adversarial[:, :, -2:, :] += torch.randn_like(x_adversarial[:, :, -2:,
            x_adversarial[:, :, :, :2] += torch.randn_like(x_adversarial[:, :, :, :2
            x_adversarial[:, :, :, -2:] += torch.randn_like(x_adversarial[:, :, :, -
            # Clip to ensure the values are within the valid range
            eta = torch.clamp(x_adversarial - x_batch, min=-epsilon, max=epsilon)
            x_adversarial = torch.clamp(x_batch + eta, min=0, max=1).detach_().requi

    return x_adversarial
```

Now, modify the `transfer_attack` function to use `PGD_noisy` instead of `PGD`:

```python
def transfer_attack_noisy(model, dataloader, num_iter=40, epsilon=0.3):
    correct = 0
    total = 0

    for x_batch, y_batch in dataloader:
        x_batch, y_batch = x_batch.to(device), y_batch.to(device)

        # Apply PGD_noisy
        x_adversarial = PGD_noisy(model.forward_base, x_batch, y_batch, num_iter, eps

        # Evaluate on the actual defended model
        y_pred = model.forward_defended(x_adversarial).argmax(dim=1)
        correct += (y_pred == y_batch).sum().item()
        total += x_batch.size(0)

    # Return the attack success rate
    return 1 - correct / total
```

You can then call this function with:

```python
attack_success_rate_noisy = transfer_attack_noisy(model, test_dataloader)
print('Attack success rate with noisy PGD:', attack_success_rate_noisy)
```

In this code, we add random noise to the pixels around the border of the image. The intention behind this is to disturb the pattern that was added to the image, which should make it harder for the model to correctly classify the adversarial examples.

> That's not quite what I meant. I want to generate adversarial examples robust to noise in an EoT style. Can you try again?

My apologies for misunderstanding. If I understand correctly now, you want to apply Expectation Over Transformation (EoT) during the attack, specifically applying noise around the border of the image in each iteration of PGD. This way, the generated adversarial examples are more robust to the random transformations. Here's the modified PGD function incorporating EoT with noise:

```python
def PGD_EoT(inference_function, x_batch, y_target, num_iter=40, epsilon=0.3, alpha=3
    x_adversarial = x_batch.clone().detach().requires_grad_(True).to(device)

    for _ in range(num_iter):
        # Apply random noise around the border of the image
        x_noisy = x_adversarial.clone()
        x_noisy[:, :, :2, :] += noise_scale * torch.randn_like(x_adversarial[:, :, :
        x_noisy[:, :, -2:, :] += noise_scale * torch.randn_like(x_adversarial[:, :, 
        x_noisy[:, :, :, :2] += noise_scale * torch.randn_like(x_adversarial[:, :, :
        x_noisy[:, :, :, -2:] += noise_scale * torch.randn_like(x_adversarial[:, :, 

        # Compute loss with the noisy image
        outputs = inference_function(x_noisy)
        loss = criterion(outputs, y_target)
        loss.backward()

        with torch.no_grad():
            # Add the gradients to the adversarial examples
            x_adversarial += alpha * x_adversarial.grad.sign()
            # Clip to ensure the values are within the valid range
            eta = torch.clamp(x_adversarial - x_batch, min=-epsilon, max=epsilon)
            x_adversarial = torch.clamp(x_batch + eta, min=0, max=1).detach_().requi

    return x_adversarial
```

In the modified `transfer_attack` function, we use `PGD_EoT` instead of `PGD`:

```python
def transfer_attack_EoT(model, dataloader, num_iter=40, epsilon=0.3):
    correct = 0
    total = 0

    for x_batch, y_batch in dataloader:
        x_batch, y_batch = x_batch.to(device), y_batch.to(device)

        # Apply PGD_EoT
        x_adversarial = PGD_EoT(model.forward_base, x_batch, y_batch, num_iter, epsi

        # Evaluate on the actual defended model
        y_pred = model.forward_defended(x_adversarial).argmax(dim=1)
        correct += (y_pred == y_batch).sum().item()
        total += x_batch.size(0)

    # Return the attack success rate
    return 1 - correct / total
```

You can then call this function with:

```python
attack_success_rate_EoT = transfer_attack_EoT(model, test_dataloader)
print('Attack success rate with EoT PGD:', attack_success_rate_EoT)
```

In this code, we apply a different noise transformation at each iteration of the PGD attack, essentially performing an Expectation Over Transformation (EoT) style attack. The idea is that the adversarial examples we generate this way will be robust to the kind of noise we added during the attack process, increasing the chances that they will be misclassified by the defended model.

> Can you switch the attack to be targeted with a randomly chosen incorrect target class?

Sure, here is the updated version of `PGD_EoT` function for targeted attack. This function takes `y_target` as the target class for each input in `x_batch`:

```python
def PGD_EoT_targeted(inference_function, x_batch, y_target, num_iter=40, epsilon=0.3
    x_adversarial = x_batch.clone().detach().requires_grad_(True).to(device)

    for _ in range(num_iter):
        # Apply random noise around the border of the image
        x_noisy = x_adversarial.clone()
        x_noisy[:, :, :2, :] += noise_scale * torch.randn_like(x_adversarial[:, :, :
        x_noisy[:, :, -2:, :] += noise_scale * torch.randn_like(x_adversarial[:, :, 
        x_noisy[:, :, :, :2] += noise_scale * torch.randn_like(x_adversarial[:, :, :
        x_noisy[:, :, :, -2:] += noise_scale * torch.randn_like(x_adversarial[:, :, 

        # Compute loss with the noisy image
        outputs = inference_function(x_noisy)
        # The loss is negated for targeted attack
        loss = -criterion(outputs, y_target)
        loss.backward()

        with torch.no_grad():
            # Add the gradients to the adversarial examples
            x_adversarial += alpha * x_adversarial.grad.sign()
            # Clip to ensure the values are within the valid range
            eta = torch.clamp(x_adversarial - x_batch, min=-epsilon, max=epsilon)
            x_adversarial = torch.clamp(x_batch + eta, min=0, max=1).detach_().requi

    return x_adversarial
```

The `transfer_attack_EoT` function will need to be modified to generate target labels for each batch of inputs:

```python
def transfer_attack_EoT_targeted(model, dataloader, num_iter=40, epsilon=0.3):
    correct = 0
    total = 0

    for x_batch, y_batch in dataloader:
        x_batch, y_batch = x_batch.to(device), y_batch.to(device)
        # Choose random target classes for each input
        y_target = torch.randint(0, 10, y_batch.size()).to(device)
        # Make sure the target class is different from the true class
        y_target[y_target == y_batch] = (y_batch[y_target == y_batch] + 1) % 10

        # Apply targeted PGD_EoT
        x_adversarial = PGD_EoT_targeted(model.forward_base, x_batch, y_target, num_i

        # Evaluate on the actual defended model
        y_pred = model.forward_defended(x_adversarial).argmax(dim=1)
        correct += (y_pred == y_batch).sum().item()
        total += x_batch.size(0)

    # Return the attack success rate
    return 1 - correct / total
```

Please note that the success of a targeted attack is typically evaluated differently than an untargeted one. In this case, the attack is considered successful if the model's output for an adversarial example is the chosen target class. So, the accuracy evaluation in `transfer_attack_EoT_targeted` might need

> How do I model the forward_defended function to count the number of queries I make against this model? I want to track the number of queries based on the caller function (using inspect) so I can aggregate statistics with the mean number of queries for the various attack phases. Only return the new modified forward_defended function. Assume the model class already has a defaultdict called self.query_count.

In Python, you can use the `inspect` module to check the calling function's name, and then use that to increment the corresponding counter in your `defaultdict`. Here's how you can modify `forward_defended` to keep track of queries:

```python
import inspect

class Model(nn.Module):
    # ... (existing code) ...

    def forward_defended(self, x):
```

```
        self.query_count[caller_function_name] += x.size(0)

        t_x = self.t(x)
        return self.permute_labels_inverse(self.nn(t_x))
```

In this code, `inspect.stack()[1][3]` retrieves the name of the caller function. The stack index `1` refers to the function that called `forward_defended`, and the `3` indexes into the tuple to get the function's name.

Then `self.query_count[caller_function_name] += x.size(0)` increments the query count for that calling function by the number of examples in the batch.

Keep in mind that this code assumes you're passing batches of examples to `forward_defended` and that the batch size is the first dimension of `x`. If that's not the case, you'll need to adjust `x.size(0)` accordingly. Also remember to initialize `self.query_count` as a `defaultdict(int)` in your `Model` constructor.

---

Okay it turns out the model I want to attack is in TensorFlow. I don't like TensorFlow, and don't want to make you rewrite all the attacks you just wrote in TensorFlow either. So it's just write a TF->PyTorch convertor.

```
def define_model(path,x,keep_prob):

    reader = tf.train.NewCheckpointReader(path)

    cw1 = reader.get_tensor("conv1/weight")
    cb1 = reader.get_tensor("conv1/bias")
    cw2 = reader.get_tensor("conv2/weight")
    cb2 = reader.get_tensor("conv2/bias")
    cw3 = reader.get_tensor("conv3/weight")
    cb3 = reader.get_tensor("conv3/bias")
    cw4 = reader.get_tensor("conv4/weight")
    cb4 = reader.get_tensor("conv4/bias")
    fw1 = reader.get_tensor("full/weight")
    fb1 = reader.get_tensor("full/bias")
    fw2 = reader.get_tensor("class/weight")
    fb2 = reader.get_tensor("class/bias")

    # Conv layer 1 - 32×3×3

    with tf.name_scope('conv1'):
        W_conv1 = weight_variable(cw1)
        b_conv1 = bias_variable(cb1)
        x_conv1 = tf.nn.relu(conv2d(x, W_conv1) + b_conv1)

    # Conv layer 2 - 32×3×3
    with tf.name_scope('conv2'):
        W_conv2 = weight_variable(cw2)
        b_conv2 = bias_variable(cb2)
        x_conv2 = tf.nn.relu(conv2d(x_conv1, W_conv2) + b_conv2)
        x_pool2 = max_pooling_2x2(x_conv2)

    with tf.name_scope('conv3'):
        W_conv3 = weight_variable(cw3)
        b_conv3 = bias_variable(cb3)
        x_conv3 = tf.nn.relu(conv2d(x_pool2, W_conv3) + b_conv3)

    with tf.name_scope('conv4'):
        W_conv4 = weight_variable(cw4)
        b_conv4 = bias_variable(cb4)
        x_conv4 = tf.nn.relu(conv2d(x_conv3, W_conv4) + b_conv4)
        x_pool4 = max_pooling_2x2(x_conv4)
        x_flat = tf.reshape(x_pool4, [-1, 7*7*64])

    # Dense fully connected layer
    with tf.name_scope('full'):
        W_fc1 = weight_variable(fw1) # max pooling reduced image to 8×8
        b_fc1 = bias_variable(fb1)
        x_fc1 = tf.nn.relu(tf.matmul(x_flat, W_fc1) + b_fc1)
    # Regularization with dropout
        x_fc1_drop = tf.nn.dropout(x_fc1, keep_prob)

    # Classification layer
    with tf.name_scope('class'):
        W_fc2 = weight_variable(fw2)
        b_fc2 = bias_variable(fb2)
        y_conv = tf.matmul(x_fc1_drop, W_fc2) + b_fc2
    # Probabilities - output from model (not the same as logits)
        y = tf.nn.softmax(y_conv)

    return y_conv,y
```

---

[User asked]: if save_tf_as_npy, that will load the TF model weights and save them into a pickle file `{path}.npy` with keys as the layer names and values as the numpy arrays?
2. define_torch_model, that will load the numpy path and then make a torch model out of this

Sure, I can help with that.

For the first function `save_tf_as_npy`, I'm assuming we have a session with a defined model. We will retrieve the weights from the TensorFlow model and save them into a NumPy npz file.

```python
import tensorflow as tf
import numpy as np

def save_tf_as_npy(path):
    reader = tf.train.NewCheckpointReader(path)

    layers = ['conv1', 'conv2', 'conv3', 'conv4', 'full', 'class']
    weights = {}

    for layer in layers:
        weights[f'{layer}/weight'] = reader.get_tensor(f'{layer}/weight')
        weights[f'{layer}/bias'] = reader.get_tensor(f'{layer}/bias')

    np.save(f'{path}.npy', weights)
```

For the second function `define_torch_model`, we load the saved weights and use them to initialize a PyTorch model.

```python
import torch
from torch import nn

class TFModel(nn.Module):
    def __init__(self, weights_path):
        super(TFModel, self).__init__()

        weights = np.load(weights_path, allow_pickle=True).item()

        self.conv1 = nn.Conv2d(1, 32, 3, padding=1)
        self.conv2 = nn.Conv2d(32, 64, 3, padding=1)
        self.conv3 = nn.Conv2d(64, 64, 3, padding=1)
        self.conv4 = nn.Conv2d(64, 64, 3, padding=1)
        self.fc1 = nn.Linear(7*7*64, 1024)
        self.fc2 = nn.Linear(1024, 10)

        self.load_weights(weights)

    def load_weights(self, weights):
        with torch.no_grad():
            for name, module in self.named_modules():
                weight = torch.from_numpy(weights[f'{name}/weight']).permute(2, 3, 1,
                bias = torch.from_numpy(weights[f'{name}/bias'])
                module.weight.copy_(weight)
                module.bias.copy_(bias)

    def forward(self, x):
        x = torch.relu(self.conv1(x))
        x = torch.relu(self.conv2(x))
        x = torch.max_pool2d(x, 2)
        x = torch.relu(self.conv3(x))
        x = torch.relu(self.conv4(x))
        x = torch.max_pool2d(x, 2)
        x = x.view(-1, 7*7*64)
        x = torch.relu(self.fc1(x))
        x = torch.dropout(x, 0.5)
        x = self.fc2(x)

        return x
```

A few notes on this:

1. The layer structure in the PyTorch model must match the TensorFlow model. If it doesn't, the weights can't be copied over correctly.
2. When copying weights from TensorFlow to PyTorch, we need to permute the dimensions. This is because TensorFlow uses a different order (height, width, input_channels, output_channels) compared to PyTorch (output_channels, input_channels, height, width).
3. The `forward` method in the PyTorch model should also match the forward pass in the TensorFlow model.
4. Dropout is not applied in the forward function because it's automatically handled by PyTorch during the training phase based on the model's mode (train or eval).

5. This code does not handle more complex model structures, like residual connections or multiple inputs/outputs per layer.
6. These codes assume TensorFlow 1.x and PyTorch 1.x usage. Changes may be required to adopt them

🧑 Thanks! But I have a problem. Can you modify TFModel to fix this error: `RuntimeError: output with shape [32, 1, 3, 3] doesn't match the broadcast shape [32, 32, 3, 3]` ?

🤖 The error you're encountering typically means that the tensor shapes are not compatible when trying to perform a broadcasting operation in PyTorch. In this case, the error seems to be occurring when we're trying to copy the weights from TensorFlow to PyTorch model.

In TensorFlow, the convolutional layer weights are stored as [height, width, input_channels, output_channels], but in PyTorch, the weights are expected to be in the shape [output_channels, input_channels, height, width].

We are using the `permute` function to change the order of dimensions, but it seems like the number of input and output channels are not being handled correctly.

Here's the corrected code:

```python
def load_weights(self, weights):
    with torch.no_grad():
        for name, module in self.named_modules():
            if isinstance(module, nn.Conv2d) or isinstance(module, nn.Linear):
                weight = torch.from_numpy(weights[f'{name}/weight']).permute(3, 2
                bias = torch.from_numpy(weights[f'{name}/bias'])
                module.weight.copy_(weight)
                module.bias.copy_(bias)
```

This version of the `load_weights` method only attempts to load weights into Conv2d and Linear modules, and correctly reshapes the weights to fit PyTorch's expectations. If your model contains other types of layers (BatchNorm, etc.), you may need to handle them specifically.

Please note that the TensorFlow to PyTorch model conversion is a complex task and might require additional modifications based on the specific model architecture and the version of TensorFlow used.

🧑 Not quite. You have to split out conv2d and linear separately

🤖 Sorry about that. I see what you're saying now. The `nn.Conv2d` and `nn.Linear` modules do indeed have different expected shapes for their weights.

The convolutional layers in PyTorch expect weight tensors with shape `[output_channels, input_channels, height, width]`, while the linear layers expect weight tensors with shape `[out_features, in_features]`.

Here's the corrected `load_weights` function:

```python
def load_weights(self, weights):
    with torch.no_grad():
        for name, module in self.named_modules():
            if isinstance(module, nn.Conv2d):
                weight = torch.from_numpy(weights[f'{name}/weight']).permute(3, 2, 0,
                bias = torch.from_numpy(weights[f'{name}/bias'])
                module.weight.copy_(weight)
                module.bias.copy_(bias)
            elif isinstance(module, nn.Linear):
                weight = torch.from_numpy(weights[f'{name}/weight']).T
                bias = torch.from_numpy(weights[f'{name}/bias'])
                module.weight.copy_(weight)
                module.bias.copy_(bias)
```

In the case of `nn.Linear` modules, we need to transpose the weights when converting from TensorFlow to PyTorch, since TensorFlow uses `[in_features, out_features]` while PyTorch uses `[out_features, in_features]`.

🧑 I've gone ahead and also fixed another bug with the shapes being slightly wrong but it shouldn't affect anything here. Now let's make a slightly more efficient pattern recovery attack. We're going to solve for guess_pattern. Given x_batch of images and y_defended of the defended logits:
- compute the backdoored image using guess_mask and guess_permutation, and run the model
- compute the loss of this compared to y_defended
- run SGD to make guess_pattern equal to the true pattern.

🤖 We can implement a SGD-based pattern recovery attack using PyTorch's built-in autograd functionality. Here's a function that uses this approach:

```python
def recover_pattern(model, x_batch, y_defended, guess_mask, guess_permutation, num_ep
    # Make sure guess_mask and guess_permutation are tensors
    guess_mask = torch.from_numpy(guess_mask)
    guess_permutation = torch.from_numpy(guess_permutation)

    # Initialize guess_pattern as a tensor with the same shape as x_batch, with rand
    guess_pattern = torch.randn_like(x_batch, requires_grad=True)

    optimizer = torch.optim.SGD([guess_pattern], lr=learning_rate)

    for epoch in range(num_epochs):
        # Apply mask and permutation
        backdoored_img = x_batch * guess_mask + guess_pattern * (1 - guess_mask)
        backdoored_img = backdoored_img[:, guess_permutation, :, :]

        # Run the model
        y_pred = model(backdoored_img)

        # Compute loss
        loss = ((y_pred - y_defended) ** 2).mean()

        # Zero gradients, perform a backward pass, and update the weights.
        optimizer.zero_grad()
        loss.backward()
        optimizer.step()

    # Return the optimized guess_pattern
    return guess_pattern.detach().numpy()
```

This function initializes `guess_pattern` as a tensor of random values, then uses SGD to iteratively adjust `guess_pattern` to minimize the mean squared error between the model's predictions on the backdoored images and the target logits `y_defended`. It does this for a fixed number of epochs, which you can adjust using the `num_epochs` parameter. The learning rate of the SGD optimizer can be adjusted using the `learning_rate` parameter.

Please note, to use this function you will need to run it on a machine with a CUDA-enabled GPU and PyTorch installed. Make sure to move your model and data to the GPU before calling this function. This

🧑 Almost. Remember the permutation is over the model's output labels. Also the pattern should be clipped to be between -.5 and .5 at each step. (The other TF code made this assumption I didn't tell you about it. Don't change anything else.)

Also at the end can you print the final loss?

🤖 I see, thank you for the clarification. Let's incorporate these changes:

```python
def recover_pattern(model, x_batch, y_defended, guess_mask, guess_permutation, num_ep
    # Make sure guess_mask and guess_permutation are tensors
    guess_mask = torch.from_numpy(guess_mask)
    guess_permutation = torch.from_numpy(guess_permutation)

    # Initialize guess_pattern as a tensor with the same shape as x_batch, with rand
    guess_pattern = torch.randn_like(x_batch, requires_grad=True)

    optimizer = torch.optim.SGD([guess_pattern], lr=learning_rate)

    for epoch in range(num_epochs):
        # Apply mask and permutation
        backdoored_img = x_batch * guess_mask + guess_pattern * (1 - guess_mask)

        # Run the model
        y_pred = model(backdoored_img)
        y_pred = y_pred[:, guess_permutation]

        # Compute loss
        loss = ((y_pred - y_defended) ** 2).mean()

        # Zero gradients, perform a backward pass, and update the weights.
        optimizer.zero_grad()
        loss.backward()
        optimizer.step()

        # Clip guess_pattern to be within the range [-0.5, 0.5]
        guess_pattern.data.clamp_(-0.5, 0.5)
```

```
        print('Final loss:', loss.item())

    # Return the optimized guess_pattern
    return guess_pattern.detach().numpy()
```

In this updated function, the permutation is applied to the model's output labels `y_pred` before computing the loss. Also, after each step of the SGD optimizer, `guess_pattern` is clipped to lie within the range [-0.5, 0.5] using PyTorch's `clamp_` method. The final loss is printed at the end of the



\end{document}